\documentstyle[preprint,aps]{revtex}
\begin{document}
% for two column  activate the line below...
%\twocolumn[\hsize\textwidth\columnwidth\hsize\csname @twocolumnfalse\endcsname
\title{Electron propagation in orientationally
disordered fullerides}
\author{E.J. Mele and S.C. Erwin}
\address{Department of Physics and Laboratory for Research on the
Structure of Matter\\ University of Pennsylvania, Philadelphia,  Pennsylvania
19104}
\date{January 24, 1994}
\maketitle
\begin{abstract}
We study the electronic spectrum for doped electronic states in the
orientationally disordered $M_3$C$_{60}$ fullerides.  Momentum-resolved
Green's functions are calculated within a cluster-Bethe-lattice model,
and compared with results from calculations on periodically repeated
supercells containing quenched orientational disorder.  Despite the
relatively strong scattering from orientational fluctuations, the
electronic states near the Fermi energy are well described by
propagating states characterized by an effective Bloch wave vector, and a
mean free path $\ell \approx$ 20 \AA.  The effective Fermi surface is
calculated in this model.  This differs from that previously calculated
for the orientationally ordered crystal, but is relatively well
described within a disorder-averaged virtual-crystal Hamiltonian, which
we derive.
\end{abstract}
\pacs{PACS numbers: 71.25.Pi, 71.25.Tn, 74.70.Jm}

% for two column  activate the line below...
%]

\section{Introduction}

Intercalation compounds formed by introducing alkali metals into the
interstitial volume of the face-centered-cubic phases of the
fullerenes have attracted a great deal of attention over the last
several years
\cite{hebard}.   The insulating undoped parent  state can be doped into a
metallic state by this method, and for the compounds $M_x$C$_{60}$ with
$x$=3 and $M$=(K, Rb) the ground state in these phases is
superconducting.  Since the first structural studies on these
compounds by X-ray diffraction \cite{steph}, it has been realized that
the orientational structure of these doped phases is quite different
from that of the undoped C$_{60}$ solid.  Undoped C$_{60}$ undergoes a
first-order phase transition from an orientationally disordered
high-temperature phase to a lower-symmetry orientationally ordered
four-sublattice structure at T $\approx 250$ K
\cite{heiney,raviharris}.  In the $x$=3 doped phase, the intensities of
the low-temperature diffraction peaks are well described by an
orientationally disordered model.  In this phase, each C$_{60}$
molecule adopts a high-symmetry orientation with its two-fold axes
aligned parallel to the [001] crystal directions.  There are
two inequivalent ways of achieving this orientation, which are related
by a $\pi/2$ rotation about any [001] axis. The intensities of
X-ray reflections are well described by a model in which this choice
of setting varies essentially randomly from site to site in the solid.

Orientational disorder has been cited as either a dominant or
contributing factor to many of the observed electronic properties of
the normal and superconducting states of these doped systems
\cite{hebard}.  The conduction electron states in these compounds are
derived from an orbitally three-fold-degenerate state of the isolated
molecule, with $t_{1u}$ symmetry.  The vector character of this state
suggests that the intermolecular hopping amplitudes can be sensitive
to the relative molecular orientations of neighboring C$_{60}$ molecules.  In
fact, the hopping amplitudes between neighboring molecules are extremely
sensitive to the molecular orientation, owing to the fact that they are
dominated by hopping of electrons between the first several closest
sites along a given intermolecular bond.  This fact has been
emphasized by Gelfand and Lu
\cite{gelfand}, and Satpathy {\it et al.} \cite{satpathy}, who noted
that because of this effect, the orientational disorder produces
off-diagonal disorder in the one-electron Hamiltonian on an energy
scale which corresponds approximately to the Fermi energy for the
doped carriers in the $ t _{1u}$ conduction band.  Some of the effects
on this disorder on the transport properties and the
frequency-dependent conductivity have since been explored
\cite{gelfb,deshpande}.

Until quite recently, theoretical work on the electronic properties of
the orientationally disordered phases has been limited to numerical
simulations on supercells containing quenched orientational disorder.
We recently reported a different approach to this problem, in which we
developed an ``effective medium" to describe the strong scattering
effects in the orientationally disordered phases \cite{deshpande}.
The theory is based on a cluster-Bethe-lattice approach in which the
disordered medium is modeled by a three-band Bethe lattice, into which
we embed a molecular cluster whose scattering properties are then
studied.  The scattering of the conduction electrons by the static
orientational fluctuations in such a cluster can be accurately studied
by this method with only modest computational effort.  In
Ref.~\cite{deshpande}, we demonstrated that this model successfully
accounts for the structure of the one-electron spectrum, and for the
low-frequency electronic excitations obtained from numerical work on
large simulation cells.

Our calculations using this method demonstrate, somewhat surprisingly,
that the effective elastic mean free path in the quenched disordered
phases must extend though several coordination shells on the fcc
lattice.  It had been previously supposed that the strong fluctuations
in the one-particle Hamiltonian would leave the merohedrally
disordered solid in an extremely strong scattering regime for which
the effective mean free path could not be meaningfully extended beyond
a nearest-neighbor distance.  This conclusion seems to have been
supported to some extent by experimental measurements of the
extrapolated residual resistivity, which have led to an estimate of
the mean free path $\ell \approx 3$ \AA.  By contrast, in
Ref.~\cite{deshpande} we found that the structure of the $t_{1u}$
conduction band is accurately described only after the effects of
phase-coherent propagation of electrons in closed rings of bonds
through at least three coordination shells are retained in the
calculations, so that one may estimate $\ell \approx 20$ \AA.  This
distance is sufficiently large that a k-space analysis of the
conduction electron states is both possible and useful, as is the case
for many other disordered alloys \cite{alloy}.  In this paper, we
expand on this observation by developing this model in k-space and
applying it to study the electron propagation in the orientationally
disordered structure.  As a point of comparison we will develop the
analogous description for both a reference orientationally ordered
doped phase, and for a model of a short-period ordered (but
orientationally modulated) model.  Our central result is that the
conduction states for the quenched disordered phases do indeed show
substantial, and quite possibly measurable, dispersion in k-space;
this structure is very different from that found for the other ordered
orientational alloys.  We do find that the dispersion in the fully
disordered phase can be successfully modeled by a
``disorder-broadened" spectrum obtained from a particular ordered
reference virtual crystal which we will describe.  Finally, we point
out that the dispersion in the single-particle spectral function might
be directly accessible by an experimental probe such as angle-resolved
photoemission.  Such an experiment would be useful since it would
provide both a spectroscopic probe of the orientational structure of
the doped phases, and a test of the validity of the underlying
one-electron picture adopted for these phases.

\section{Effective Medium Theory}

In this section we provide a brief review of the Bethe-lattice
construction which was originally reported in Ref.~\cite{deshpande}.

As noted in the Introduction, the conduction states in the alkali
fullerides are derived from the orbitally three-fold-degenerate
$t_{1u}$ state of the isolated molecules.  The broadening of this
level into a conduction band can be described by introducing a
nearest-neighbor 3$\times$3 matrix hopping amplitude, $T _{\mu
\nu} (\tau)$, describing the hopping  amplitudes between states of orbital
polarizations $\mu$ and $\nu$ across the bond $\tau$.  The matrix
elements in $T$ depend on the orientation of the bond $\tau$ and on the
orbital polarizations.  In their original work on this problem,
Gelfand and Lu (GL) noticed that the signs and sizes of the various
elements in $T$ depend on the relative molecular orientations on the
terminal sites \cite{gelfand}.  Their results for the matrix hopping
amplitudes are reproduced in Table I.

The calculations of GL are carried out using a particular convention
for labeling the orbital polarizations on each site, in which the
coordinate axes are ``fixed" in the frame of the molecule, and
therefore spatially varying in the crystal frame (for a discussion of
this point see Ref.~\cite{yildirim}).  It is useful to reformulate this
result in a way which separates the effects of rotation of the
coordinate system from the effect of variation of the underlying
hopping amplitudes.  A simple method is to construct a matrix
amplitude $H _{bond}$ which describes the effects of forward and
backward propagation across the bond $\tau$ in the manner
\begin{equation} H_{bond}(\tau) = \left( \begin{array} {cc} 0&T_{M,N}
\\ T^T_{M,N} & 0
\end{array} \right) \end{equation}
Here $M$ and $N$ can take the values ``$A$" and ``$B$" denoting the
orientation of the two sites connected by the hop. Squaring $H_{
bond}$ in Eq.~(1) decouples the two sites, demonstrating that the
eigenvalues of Eq.~(1) must be ordered in pairs $\pm \mid t_n \mid$, for
any choice of the underlying matrix $T$.  Each six-component
eigenvector, $V$, of $H_{ bond}$ can be then built up from two
normalized three-component vectors $[u_n]$ ($n$=$i$,$j$) in the manner $V =
(u_i,u_j)$.  For an eigenvector of the bond-hopping operator in the
form ($u_i , u_j$) with eigenvalue $e_n$, the vector $V' = (u_i ,
-u_j)$ is an eigenvector with eigenvalue $ -e_n$. Thus the action of a
general hopping matrix $T$ can always be reduced to the elementary 3$\times$3
matrix operation $T = U_j ^T h U_i$, where $h$ is a diagonal matrix
containing (by convention) only the $ positive $ eigenvalues of
$H_{bond}$ in Eq.~(1), and $U_i$ is a matrix whose rows are the three
components of the associated $V$'s projected onto the $i$th site.  Here,
the action of the internal hopping operator is positive definite, and
all the sign and polarization conventions are contained in the $U$
rotation matrices.

The nonzero elements in the residual diagonal matrix $h$ can still
depend on the relative molecular orientations.  However, by a direct
calculation of these, one finds that these nonzero diagonal matrix
elements depend only weakly on the molecular orientation.  As an
example, we list in Table II the positive eigenvalues of $H_{bond}$
constructed for a hop between a bond between two molecules of like
orientation (denoted $AA$), and two molecules of unlike orientation
(denoted $AB$).  For the most important, that is the largest of these,
the fractional variation is only of order 10\%, indicating that the
dominant effects of orientational disorder are introduced through the
$U$ matrices and not through the bare amplitudes in $h$.  The very slight
variation of these eigenvalues in the two configurations occurs
because in both the $AA$ or $AB$ orientations, a single bond is
eclipsed along the nearest-neighbor direction.  This single bond is
labeled the ``5-6" bond since it borders a fivefold and sixfold ring
on the surface of the fullerene.  In the $AA$ orientation the
fivefold and sixfold rings on neighboring molecules are eclipsed,
while in the $AB$ orientation the fivefold ring eclipses a sixfold
ring. In the construction of the effective medium model, we will
therefore ignore the variation of the eigenvalues $h$ between the two
structures and focus entirely on the very important orientation
dependence of the rotation matrices $U$.

The effect of the rotations $U$ is to induce a reorientation of the
orbital polarization of a given electron as it hops from molecule to
molecule in the structure.  If one regards the electron as a particle
which carries with it an $\ell = 1$ internal orbital degree of
freedom, this Hamiltonian then describes a type of spin-orbit
interaction coupling the orbital motion on the lattice to the internal
orbital polarization of the particle. However, we emphasize that in
this problem we are in a very strong scattering limit, where the
reorientation of the orbital polarization is in no sense a ``weak"
precession of its internal moment as the particle propagates from
site to site on the lattice.

The total density of states can be obtained from a trace of the
one-particle Green's function $G_+(E) = \left( E + i \delta - H \right
) ^{-1}$, where the trace requires a sum over sites and over orbital
polarizations on a given site.  If one develops an expansion of the
diagonal elements of $G$ in powers of $(1/E)H$:
\begin{eqnarray}
G(E) &=& \left( \frac{1}{E} \right) + \left( \frac{1}{E} \right)
^2 H + \left( \frac{1}{E} \right) ^3 H H \nonumber \\
&& + \left( \frac{1}{E} \right) ^4 H H H + etc.,
\end{eqnarray}
each term appearing in tr$G$ can be interpreted as a specific closed
electron trajectory on the lattice.  For propagation in a disordered
structure in which the $U$'s are varying essentially randomly from
bond to bond in the structure, one expects that large closed loops
will not contribute spectral weight in the one-particle Green's
function, since any reference orbital polarization will be scattered
symmetrically into all possible orientations after its propagation on
a closed loop.  This cancellation rule fails for an orientationally
ordered crystal since the large $N$-fold loops remain highly
correlated for large $N$.  Even in the orientationally disordered
system, the special class of Brinkman-Rice \cite{brinkman} retraceable
paths, which propagate randomly from some reference site and exactly
retrace their steps back to the origin, always makes a
positive-definite contribution to the trace for large $N$, and
ultimately determines the long-distance behavior of the Green's
function.

In view of this we are motivated to study the properties of an
effective medium in which only the retracing paths are retained in the
Green's function.  For our problem, this network has the topology of a
tree with one ``ingoing" and eleven ``outgoing" bonds at every node.
The hopping amplitude along any single nearest neighbor bond on this
network is given by the 3$\times$3 $T$ matrices discussed above, which
explicitly depend on the orientation of the bond.  In the
approximation that we neglect the weak dependence of the matrix $h$ on
molecular orientation (see Table II), the intermolecular hopping
amplitudes between like ($AA$ or $BB$) and unlike ($AB$ or $BA$) orientations
may be locally transformed into each other by suitable local rotations
of the orbital bases.  Of course, on a physical lattice this freedom
will be eliminated by the closure of rings of bonds.  On the tree,
which contains no closed rings, we will consider the dynamics on a
reference structure in which each site is arbitrarily assigned the
``$A$'' orientation.  Thus every node of the tree is equivalent, and one
can then exploit this symmetry to construct the Green's function for
the medium.

The construction of the Green's function on the tree is straightforward.  The
solution  can be compactly expressed in terms of a complex matrix effective
field $\Phi (E, \tau)$  which depends on energy $E$ and the bond orientation
$\tau$ such that:
\begin{equation}
G_{j,k} (E)  = \Phi({\tau_{ij}};E) G_{i,k} (E),
\end{equation}
where the subscripts denote lattice sites, and site $j$ is a nearest
neighbor to site $i$. The Dyson equation for $ EG(E) = I + HG $ provides
a self-consistency condition for the field $\Phi$:
\begin{equation} E \Phi(\tau)  = H (\tau) + \sum _{\tau \neq \tau'} H  (\tau')
\Phi (\tau') \Phi (\tau),
\end{equation}
from which $\Phi $ can be determined. In terms of the fields $\Phi$,
the diagonal block of the one-particle Green function at any site in
the infinite tree is:
\begin{equation} G_{0,0} = [E -
\sum_{\tau}  H (\tau) \Phi (\tau, E)]^{-1}.
\end{equation}
The medium we construct in this manner will ultimately be used to
provide a boundary condition for studying electron motion within
(relatively small) orientationally disordered clusters extracted from
the full orientationally disordered fcc lattice.  Before turning to
that application however, it is useful to examine the spectrum of the
isolated tree, for which the density of states is shown in
Fig.~\ref{bldos}(a).  (Here and elsewhere in the paper, all energies
are expressed in units of $t$, which sets the energy scale for the matrix
elements listed in Table I. Estimates of $t$ range between 10 meV and
14 meV for the doped systems.)  We observe that this spectrum for the
medium is an even function of energy, because all the
retracing paths appearing in $G$ require exactly an even number of
steps. The overall bandwidth obtained in this calculation is approximately
36$t$, which is very nearly the bandwidth obtained from direct
numerical calculations on orientationally disordered supercells.  This
confirms the original conjecture that it is the retraceable paths
which control the analytic structure in the one-particle Green's
function for the disordered system.  Finally, the spectrum is smooth
and relatively structureless, reflecting the absence of closed
coordination rings on the network.  In fact, although the overall
bandwidth agrees relatively well with that obtained from studies on
disordered cells, the spectrum obtained for the tree is noticeably
more symmetric and less structured than the spectra obtained from the
numerical simulations on disordered lattices
\cite{deshpande}.  The structure obtained in the densities of state
for the disordered lattices can be traced to the effects of the
closure of the rings of bonds on the fcc network; these effects
will be revealed more clearly by the momentum-space analysis given
below.

We should remark that while the full matrix self-energy $\Sigma (E) =
\sum_{i} H_i \Phi (\tau_i)$ is a complex $\it{diagonal}$ 3$\times$3 matrix
(in orbital space) and therefore does not mix orbital polarizations,
each of the individual contributions $\sigma_i = H_i \Phi (\tau_i)$ does
have important nonvanishing off-diagonal components.  These terms
describe contributions to the on-site self-energy from indirect
processes in which a particle, starting from a reference site,
propagates into the tree and undergoes a reorientation of its orbital
polarization before its return.  The off-diagonal terms thus describe
an orbital dephasing of the electron by its coupling to the external
medium.

We now consider the effect of embedding a finite molecular cluster
embedded in the disordered medium treated in this tree approximation.
The full Hamiltonian is partitioned:

\begin{equation} H = H_c  + H_m + H_{cm} \end{equation}
where $H_c$ contains terms connecting two sites inside the cluster,
$H_m$ refers to bonds connecting neighboring sites in the effective medium,
and the $H_{cm}$ represents the connecting bonds at the boundaries of the
cluster. The ``internal" Hamiltonian $H_c$ is thus treated exactly in
this formulation, and in particular the closure of coordination rings
is completely accounted for within the molecular cluster in $H_c$.
The Green's function for the composite medium can be projected onto
this cluster by the matrix inversion:
\begin{equation}  G(E) = \left( E - H_c - \sum_{i} \psi_i^{\dagger}
\left( \sum_{j(i)} \sigma_{j} \right) \psi_i\right) ^{-1},
\end{equation}
where the sum is over the boundary sites $i$ connected through the bonds
$j$ to the external medium.  Thus the effects of indirect propagation in
the medium appear through the self-energy contributions $\sigma_j$.

\section{Densities of States}
In this Section, we provide an analysis of the Green's function of
Eq.~(7) in momentum space.  Before turning to those results, it is
useful to examine the k-integrated total densities of states obtained
for these systems.

In Fig.~\ref{bldos}(b) we show the densities of states projected onto a single
tetrahedral prism extracted from the fcc structure (the prism consists
of four sites that are mutually nearest neighbors), embedded in the
effective medium described above.  Results are shown for the three
inequivalent molecular configurations $A_4$, $A_3B$ and $A_2B_2$
within this tetrahedral cluster.  One finds that the closure of the
threefold rings of bonds perturbs the spectral structure obtained for
the isolated tree.  For the two configurations which contain bonds
between molecules with inequivalent orientations, the effect is to
provide a weak enhancement of the spectral weight at negative energy
in the conduction band.  However, it is also clear from the broad and
relatively structureless character of the spectra that
the electronic states in these clusters are dominated by the orbital
dephasing introduced through coupling to the effective medium.

Similar calculations, in which the size and symmetry of the embedded
cluster are varied, demonstrate that due to the large coordination
number on the fcc lattice, the dephasing of the orbital polarization
introduced by the effective medium is a very strong effect in this
problem.  Consequently, one must develop this model on relatively
large embedded clusters before the data converge suitably to those
obtained from numerical simulations on fully orientationally
disordered lattices.  As an example we reproduce, in Fig.~\ref{bldos}(c), data
previously presented in Ref.~\cite{deshpande} for embedded 19- and
43-site molecular clusters.  These models just close the (002) and
(112) coordination shells on the fcc lattice, respectively.  Here we
find that after closing the third coordination shell the spectra do
provide an accurate representation of the spectra obtained from
quenched orientationally disordered lattices (numerical results
obtained from an average over an ensemble of 27-site lattices are
shown for comparison). From this comparison one concludes that the
single-particle Green's function is sensitive to phase-coherent
propagation of the electrons over a range of order $\ell \approx $ 20
\AA.  Although relatively small by absolute standards, this length is
surprisingly large given the observation that the scale of fluctuating
term in the Hamiltonian for the $t_{1u}$ electrons in the
orientationally ordered phase is the one-electron bandwidth.  However,
the result can be rationalized by noting that these fluctuations
appear only in the off-diagonal terms in the Hamiltonian, and can
therefore be locally removed in any single bond by a gauge
transformation to a suitable choice of orbital bases on two
neighboring sites.  To study the residual frustration to the
electron motion we need to examine the propagation of the electrons on
closed loops in the network; this can be described by a
gauge-invariant correlation function.

\section{Momentum Space Formulation}

In this section we reanalyze the results of Section III by formulating
the theory in momentum space. This will be useful for understanding the
propagation and the orbital structure of the conduction electron
states, and for interpreting the various static-response functions
which are mediated by the conduction electrons in the disordered
phase.

Formally, our approach involves a simple rotation of the Green's
function of Eq.~(7):
\begin{equation} G(k,E) = \sum_{i,j} R^{\dagger}_i(k) G_{i,j} (E) R_j(k)
\end{equation}
with $R_i(k) = \exp (i k \cdot T_i)$ where $T_i$ is a lattice site
locating the $i$th molecular state.  One observes that for a
hypothesized model in the strong-disorder limit, for which the
off-diagonal elements in the Green's function decay very quickly on
the scale of a lattice spacing, only the diagonal elements of $G$ can
contribute appreciable weight in the sum in Eq.~(8), and therefore
the projected $G$ is therefore approximately k-independent.  The results
of Section II of course indicate that the nonlocal terms in $G$ do
contribute significantly even in our fully disordered model; the
resulting dispersion of the spectral function $A(k,E) = (1/\pi)
{\rm Im} G(k,E)$ will be studied in this section. The calculated density
$A(k,E)$ gives the probability that an electron at energy $E$ occupies a
state indexed by crystal momentum $k$.

In Fig.~\ref{dis} we plot the spectral function $A(k,E)$ calculated
for wave vectors along the $\Gamma - X$ $ \langle 001 \rangle$ and
$\Gamma -L$ $\langle 111 \rangle$ symmetry directions in reciprocal
space, referred to the Brillouin zone of the reference fcc structure.
These data are obtained carrying out the projection of Eq.~(8) over a
molecular cluster containing 79 sites.  This closes the (310) shell
around a reference site at the origin, with the boundary sites
terminated by coupling to the Bethe lattice as described earlier.
These data are presented as a surface plot of $A(k,E)$ with $k$
plotted along one horizontal axis and the energy plotted along the
orthogonal horizontal axis.  In this plot, the states at negative
energies are occupied when the conduction band is half filled.

We find that the spectral function calculated in the model is
dispersive, but broadened due to the scattering from the static
orientational fluctuations.  At the zone center, the spectral function
is strongly peaked at an energy $E = -15.4t$ relative to the Fermi
energy.  Along the $\Delta$ $\langle 001 \rangle$ direction this
feature exhibits positive dispersion, so that at the zone boundary the
spectral weight is evenly divided between positive- and
negative-energy states.
(Actually, a closer analysis reveals that this
spectrum consists of two rather well resolved components, as shown in the
inset of Fig.~\ref{dis}.  For
dispersion along the $\langle 001 \rangle$ directions the
$z$-polarized component of the $t_{1u}$ band is essentially
dispersionless, while the $x$ and $y$ components exhibit identical
positive cosine-like dispersion. Along the $\Lambda $ $\langle 111
\rangle$ direction, these three orbital
polarizations cannot be distinguished by their symmetry, and exhibit nearly
identical positive dispersion away from the zone center.  As we
explain in detail below, all of these behaviors are predicted from the
dispersion of an Hamiltonian for an orientationally averaged but $\it
{ordered}$ virtual crystal.)

For comparison, we have carried out the same calculation for an
orientationally ordered reference cluster, results for which are
shown in Fig.~\ref{ord}.  The spectral function exhibits
a dispersion which is by now quite familiar from the theory of
$M_3$C$_{60}$ in the fully orientationally ordered
structure.  The conduction band at $\Gamma$ shows a three-fold
degenerate peak at a positive energy $E = 8.2t$ above the Fermi
energy.  This peak is split into an orbital doublet and singlet as one
proceeds along the $\Lambda$ direction. Along the $\Delta$ direction,
the $t_{1u}$ multiplet is split into three separate states, two of
which exhibit negative dispersion, and one of which disperses to
higher positive energies.  Since only the states at negative energies
are occupied in the model, the band structure shown in Fig.~\ref{ord}
leads to a Fermi surface consisting of two interpenetrating sheets,
corresponding to the two bands which cross $E=0$. The topology of this
unusual Fermi surface is discussed in more detail in
Refs.~\cite{erwin} and \cite{meleerwin}.

The widths of the spectral peaks in Fig.~\ref{ord} are due to the
finite size of the molecular cluster in which the calculations are
carried out.  Comparing the spectra in Figs.~\ref{dis} and \ref{ord}
we observe that the widths for the disordered phase are significantly
larger than those obtained for the ordered model, and are therefore
not ``resolution limited", but are intrinsic to the orientational alloy.

Finally, we have carried out a similar calculation for the layered
bidirectional structure, in which alternate layers are assigned
uniformly to the $A$ and $B$ orientation, modulated along the $[001]$
direction.  The k-resolved spectral functions for this model are shown
in Fig.~\ref{bi}.  (The axis conventions are the same as adopted in
Figs.~\ref{dis} and \ref{ord}, but the Brillouin zone is now simple
tetragonal.)  Here we see more complex behavior, intermediate
between that of the models presented in Figs.~\ref{dis} and \ref{ord}.
The peaks in the spectral density are relatively sharp and strongly
dispersive.  The overall dispersion is positive, with a filled
degenerate state at the zone center which then disperses above the
Fermi energy as one approaches the Brillouin-zone boundaries.  The
results obtained here agree in detail with the dispersion obtained by
directly calculating the energies of the Bloch eigenfunctions
propagating on a periodically repeated $A_2B_2$ crystal in the tetragonal
structure. Note that the data presented in  Fig.~\ref{bi} use the conventions
for the fcc primitive cell; that is they are plotted in an extended zone
convention in which the six bands of the tetragonal crystal are unfolded into
two groups of three which disperse along the $k_z$ direction.

\section{Virtual Crystal Hamiltonian}

The spectral function calculated for the orientationally disordered
solid can be understood by studying the spectrum of a particular
ordered reference crystal.  The construction of this virtual crystal
model as a reference Hamiltonian has been previously described in
Ref.~\cite{yildirim}.  There the difference between the exact
Hamiltonian in a particular orientational state and the virtual
crystal was treated as a perturbation with which we were able to
investigate the indirect orientation-dependent interactions between
fullerenes mediated by the response of the conduction electrons.  Here
we briefly review that construction and refer the reader to
Ref.~\cite{yildirim} for a more detailed discussion.

We observe that the orientation dependence of the nearest-neighbor
matrix hopping amplitudes can be represented in the form
\begin{equation}
T_{MN} = t_{\mu, \nu} (\tau, \sigma_M, \sigma_N),
\end{equation}
where $\mu$ and $\nu$ run over the symmetry labels $x$, $y$, and $z$, and
the $\sigma$ are Ising spin variables, where $+1$ denotes the $A$
orientation and $-1$ the $B$ orientation. The four possible hopping
matrices between neighboring sites across bond $\tau$ can be neatly
decomposed into an ``average" contribution, two terms which are linear
in the spin variables, and one term which is bilinear in the spin
variables:
\begin{equation} t_{\mu,\nu} (\tau, \sigma_M, \sigma_N) = t^{(0)}_{\mu,\nu} +
t^{(1)}_{\mu,\nu}\sigma_M + t^{(2)}_{\mu,\nu} \sigma_N +
t^{(3)}_{\mu,\nu} \sigma_M
\sigma_N. \end{equation}
The virtual crystal is obtained by treating the $\sigma$ as
uncorrelated random variables with $\langle \sigma \rangle$ = 0, so
that the only nonvanishing term is $t^{0}$ in Eq.~(10).  This residual
piece can be interpreted as resulting from an ensemble average over
orientations, $\it {before}$ relaxing the conduction electron states.
Of course, since the electronic bandwidth is much larger than the
frequency scale for orientational fluctuations, one should in
principle reverse these two steps.  However, we now show that the
residual orientationally averaged ordered Hamiltonian successfully
accounts for the symmetry and structure of the spectral functions
shown in Fig.~\ref{dis} for the orientationally disordered model.

For a hop along a (110) bond direction, symmetry considerations
require that the virtual crystal $t^{(0)}$ takes the form
\begin{equation} t^{(0)} = \left( \begin{array} {ccc} A&B&0 \\ B&A&0 \\ 0&0&C
\end{array} \right).
\end{equation}
The values of $A$, $B$ and $C$ extracted from the GL Hamiltonian are given
in Table III.  Here we see that $A$ and $B$, which describe the
hopping between $t_{1u}$ orbitals which project along the bond
direction $\tau$ (that is, the $x$ and $y$ polarizations) are quite small.
This results from a near cancellation of hopping amplitudes which
change sign as one flips the molecular orientations.  The remaining
constant, $C$, represents the hopping between $z$-polarized $t_{1u}$
orbitals along the [001] direction, and dominates the intermolecular
hopping amplitude.  The analogous hopping matrices
for motion along any of the bonds of the fcc structure can be obtained
by a straightforward rotation of $t^{(0)}$ in Eq.~(11).

The orientationally averaged matrix elements in Table III are
interesting, since they suggest that it is useful to consider a
simpler model in which we drop the small $x$ and $y$ hopping amplitudes,
and consider the motion in the $xy$ plane, which is dominated by the
propagation of a $z$-polarized conduction electron.  (Similar
arguments apply for propagation in the $yz$ and $xz$ planes.) Thus if we
set the small amplitudes $A$ and $B$ listed in Table III to zero, we
recover a reference Hamiltonian in which the orbital polarizations
are completely decoupled in the model.  The dispersion relations
for these three bands are then given by the three independent spectra:
\begin{eqnarray}
2 (cos ( q_y a) + cos (q_z a) ), \makebox[.25in] & x{\rm -polarized} \nonumber
\\
2 (cos ( q_x a) + cos (q_z a) ), \makebox[.25in] & y{\rm -polarized} \nonumber
\\
2 (cos ( q_x a) + cos (q_y a) ), \makebox[.25in] & z{\rm -polarized}
\end{eqnarray}
Eqs.~(12) thus describe three interpenetrating dispersion
relations for the $x$-, $y$- and $z$-polarized electrons propagating on three
mutually orthogonal two-dimensional lattices.  If we now reintroduce
the smaller but nonvanishing elements $A$ and $B$, one finds a slight admixing
of these branches, particularly along the $\langle 111 \rangle$
directions, which induces a perturbation to this spectrum that is
relatively well confined to E = $ \pm 2t_{\|} $ around the Fermi
energy.  The density of states for this virtual crystal is shown in
Fig.~\ref{virtdos}, which clearly shows its underlying two-dimensional
character.

The virtual-crystal approximation provides a surprisingly accurate
description of the spectra calculated for the embedded clusters with
quenched orientational disorder that were presented in Section III.
The spectra shown there represent the spectral function for the
disorder-averaged Green's function $\langle G \rangle$, while in the
virtual crystal the ensemble average is applied first to the
Hamiltonian, giving $G_{vc} = [E - \langle H \rangle]^{-1}$.  Although
these are clearly very different objects, the symmetry and structure
of the one-particle states in the disordered model are
represented reliably in the virtual-crystal approximation.

The virtual crystal also correctly describes
orbitally projected spectral densities of states obtained from the
quenched average.  For example, it is clear from Eq.~(12) that the $z$
and $xy$ polarizations are decoupled for dispersion along the $\langle
001 \rangle$ direction in k-space (and our observations above suggest
that this should survive the reintroduction of the small amplitudes
$A$ and $B$).  In Fig.~\ref{proj}, the various orbital projections of
the single-particle Green's function from the embedded cluster are
displayed; they clearly exhibit exactly this behavior.  The $z$-polarized
component of the spectrum is nearly dispersionless, and is characterized
by a prominent feature at the bottom of the $t_{1u}$ conduction band.
The $xy$-polarized components clearly exhibit a cosine-like dispersion
along the $\Delta$ direction.

Finally, we turn to spectral densities at the fixed energy $E=E_f$,
which is the alloy generalization of the crystalline Fermi surface.
In Fig.~\ref{padutch} we show first the Fermi surface calculated for
the orientationally ordered crystal (upper left), and the Fermi
surface of the virtual-crystal model (upper right). For the virtual
crystal, the Fermi surface can be described as intersecting two-dimensional
``sheets".  These are the extensions into three dimensions of the
well-known inscribed square Fermi surfaces predicted by the
square-lattice spectra of Eq.~(12).  The slight warping of these
surfaces shown in the figure result from the weak nonzero elements $A$
and $B$ in the longitudinal block of $t^{(0)}$.  By comparison we
show, in the lower two panels of Fig.~\ref{padutch}, the spectral
function at fixed energy, $A(k,E_f)$, calculated in the
effective-medium theory  (lower left), and a direct
numerical construction of $A(k,E_f)$ from Fourier analysis of the
single-particle states at the Fermi energy, averaged over an ensemble
of periodically repeated 27-site supercells with quenched
orientational disorder (lower right).  The structure of the
virtual-crystal Fermi surface is clearly apparent in both of these
plots.  Moreover, the  alloy Fermi-level momentum distribution
can be described quite accurately by a disorder-broadened
virtual crystal with an effective correlation length $ \xi
\approx $ 20 \AA.  This length is completely consistent with our analysis
of the correlation length on the disordered fcc network given
in Section III.

The accuracy of the virtual crystal theory can be understood by
noticing that the resulting Fermi surface encloses the correct volume
in phase space and exhibits the same ($O_h$) symmetry as the full
disorder-averaged Green's function.  However, we should point out that
while the virtual crystal correctly predicts the symmetry of the full
spectral function, it does a poor job of quantitatively predicting the
bandwidth.  In fact, the bandwidth of the conduction band for the
quenched disordered state is nearly a factor of two larger than
predicted in the virtual crystal. These are compared in
Fig.~\ref{virtdos}, as the solid (virtual crystal) and dashed
(disordered crystal) spectra. Thus we observe that the fluctuating
terms in the Hamiltonian do play a crucial role in fixing the scale of
the intermolecular hopping amplitudes, but that the symmetries of the
resulting spectrum are quite faithfully described by the annealed
average in the simpler virtual-crystal approximation.

\section{Discussion}

Orientational disorder in the fullerides introduces off-diagonal
disorder in the effective electronic Hamiltonian for these alloys.
Although the energy scale for this disorder is relatively large, we
find that the scattering effects can be quite subtle.  In particular,
the orientation dependence of the interatomic hopping amplitudes can
be essentially removed locally in any bond, by a suitable local
rotation of the electronic basis states in the $t_{1u}$ subspace. The
residual effects of disorder then occur because the fluctuations in
the Hamiltonian cannot be consistently removed globally in every bond
by this procedure. It appears natural to formulate this problem as a
gauge model on the fcc lattice, and indeed the moment expansion in
Eq.~(2) provides just such a formulation.  Here we find that
off-diagonal correlations in the one-particle Green's function are
surprisingly robust, so that the states near the Fermi energy can be
described as propagating states on an effective Fermi surface.
Interestingly, this is different from what one obtains for the
orientationally ordered crystalline phase.

At present, direct experimental data supporting the merohedral
disorder hypothesis is obtained from X-ray scattering data. The idea
is supported indirectly by the relatively large low-temperature
resistivity, and by the observed breakdown of various selection rules
for the excitation of the intramolecular vibrational modes.  The
dispersion of the spectral functions obtained for the disordered
models studied in Section IV provide a direct microscopic probe of
the scattering effects due to static orientational disorder.  The
overall dispersion of these bands is of order 0.5 eV, and might be
directly resolvable by angle-resolved photoelectron spectroscopy.
Indeed, the widths of these features in the spectral densities are
similar to what one obtains in experiments on ordered simple metals.
Since the $\it {sign}$ of the dispersion is reversed in the ordered
and disordered models, experimental observation of the dispersion of
the single-particle spectral function can be used to test these models.
Experimental observation of this dispersion might provide a useful
test of the underlying one-electron picture which is commonly adopted
to describe these phases.  Of course this treatment ignores the effects
of coupling to the vibrational degrees of freedom, as well as the effects
of direct electron-electron interactions, which can also contribute to the
observed single-particle spectra.

The conduction electrons in these systems also contribute to the
effective potentials for the intramolecular vibrations in these doped
solids. These are difficult to calculate accurately for the
orientationally disordered systems.  The most complete studies of
these effects have made use of the supercell approximation in which
the response functions are computed for a large, periodically repeated
cell which contains orientational disorder on the fullerene sites.
The effective-medium construction presented above can be
straightforwardly applied to these studies, and provides an alternative
method for investigating these effects.

Acknowledgements: This work was supported by the NSF under the  grant  DMR
91-20668 and by the Department of Energy under grant  91ER45118.

%
% tables
%
\begin{table}
\caption{The intermolecular hopping matrices
along a (110) bond direction for molecules in like
($AA$) and unlike ($AB$) orientations.  The data are reproduced from
reference \protect\cite{gelfand}.  The orbital polarizations are defined in
the frame of the molecule, and differ by $\pi/2$ rotations for
the orientationally inequivalent case. The amplitudes are scaled by a
factor t $\approx$ 14 meV. }
\begin{tabular}{ccccccc}
& &$AA$& & \makebox[.1in] &$AB$&  \\
 &$x$&$y$&$z$&$x$&$y$&$z$ \\
\tableline
$x$&0.83&-1.98&0.00&1.75&2.08&0.00\\
$y$&-1.98&3.36&0.00&-2.08&-3.71&0.00\\
$z$&0.00&0.00&-1.91&0.00&0.00&-2.67\\
\end{tabular}
\end{table}
\begin{table}
\caption{The three positive eigenvalues of the bond
hopping operator $H_{bond}$, for transitions between molecules with like ($AA$)
and unlike ($AB$) settings. The eigenvalues of $H_{bond}$ are ordered in pairs
$\pm \mid h \mid$.  }
\begin{tabular}{cc}
$AA$& $AB$  \\
\tableline
4.445 & 5.029 \\
1.910 & 2.670  \\
0.255 & 0.431  \\
\end{tabular}
\end{table}
\begin{table}
\caption{ The three matrix elements which appear in the
orientationally averaged virtual crystal defined by Eq.~(11).}
\begin{tabular}{ccc}
$A$&$B$&$C$ \\
\tableline
0.01&0.38&-2.29 \\
\end{tabular}
\end{table}
%
% figures
%
\begin{figure}
%\epsfxsize=3in
%\centerline{\epsffile{propagation.fig1.ps}}
\caption{(a) The density of states calculated for the isolated fcc
fulleride tree. Here and elsewhere in the paper the energy axis is
scaled in units of $t$ which we estimate as $t \approx $ 14 meV.  (b)
The density of states traced over a tetrahedral cluster embedded in
the tree.  Three different orientational structures for the cluster
are compared: $A_4$, $A_3B$, and $A_2B_2$.  (c) The local density of
states at the core of 19-site (dashed) and 43-site clusters (heavy
solid) embedded in and coupled to the effective medium.  The settings
are chosen randomly in these clusters.  Also shown (light solid) is
the result of a quenched average over a series of 27-site
orientationally disordered structures.
\label {bldos}}
\end{figure}

\begin{figure}
%\epsfxsize=3in
%\centerline{\epsffile{propagation.fig2.ps}}
\caption{Spectral function $A(k,E)$ for an orientationally disordered
79-molecule cluster coupled to the effective medium. The spectra are
plotted with $k$ varying along the symmetry direction $L - \Gamma - X$
in the fcc zone. Inset: band structure for tight-binding Bloch states
of the virtual-crystal Hamiltonian of Eq.~(11).
\label {dis}}
\end{figure}

\begin{figure}
%\epsfxsize=3in
%\centerline{\epsffile{propagation.fig3.ps}}
\caption{Spectral function $A(k,E)$ for an orientationally ordered
79-molecule cluster coupled to the effective medium. Plotting
conventions are the same as adopted in Fig.~\protect\ref{dis}. Inset:
band structure for tight-binding Bloch states of the orientationally
ordered fcc crystal.
\label {ord}}
\end{figure}

\begin{figure}
%\epsfxsize=3in
%\centerline{\epsffile{propagation.fig4.ps}}
\caption{Spectral function $A(k,E)$ for an ordered bidirectional
cluster coupled to the effective medium. The plotting conventions are
the same as adopted in Figs.~\protect\ref{dis} and \protect\ref{ord}.
Inset: band structure for tight-binding Bloch states of the
bidirectional structure.
\label {bi}}
\end{figure}

\begin{figure}
%\epsfxsize=3in
%\centerline{\epsffile{propagation.fig5.ps}}
\caption{The k-integrated density of states calculated for the
orientationally averaged virtual crystal (solid) as described in the text.
The dispersion is essentially two dimensional for each orbital
polarization, with weak mixing between orbital polarizations. The density
of states for the disordered crystal is reproduced from
Fig.~\protect\ref{bldos} as a dashed curve.
\label{virtdos}}
\end{figure}

\begin{figure}
%\epsfxsize=3in
%\centerline{\epsffile{propagation.fig6.ps}}
\caption{Orbitally projected spectral functions for the disordered
system, dispersing along the $\langle 001\rangle$ directions in
k-space.  (Top) The $z$-polarized component is nearly dispersionless
and peaked near the bottom of the conduction band. (Bottom) The
$x$- and $y$-polarizations show identical cosine-like dispersion, as found in
the virtual-crystal approximation.
\label {proj}}
\end{figure}

\begin{figure}
%\epsfxsize=3in
%\centerline{\epsffile{propagation.fig7.ps}}
\caption{Comparison of the Fermi-level momentum distributions
$A(k,E_f)$ in the
orientationally ordered state (upper left); the virtual-crystal
Hamiltonian (upper right);
the disordered state as described by
the cluster-Bethe lattice model (lower left); and numerical simulations on
disordered supercells (lower right).  Note that the
distributions for the disordered models are well described by a
disorder-broadened version of the virtual crystal.
\label{padutch}}
\end{figure}

\end{document}